\documentclass[longauth, traditabstract]{aa} 

\usepackage{graphicx}
\usepackage{txfonts}
\usepackage{natbib}
\begin{document}

\title{Herschel/HIFI measurements of the ortho/para ratio in water
  towards Sagittarius~B2(M) and W31C\thanks{Herschel is an ESA space observatory with
    science instruments provided by European-led Principal
    Investigator consortia and with important participation from
    NASA.}}

\author{
D.~C.~Lis \inst{1},
T.~G.~Phillips \inst{1},
P.~F.~Goldsmith \inst{13},
D.~A.~Neufeld \inst{3}, 
E.~Herbst \inst{14}, 
C.~Comito \inst{8},  
P.~Schilke \inst{8,12}, 
H.~S.~P.~M\"{u}ller \inst{12}, 
E.~A.~Bergin \inst{2}, 
M.~Gerin \inst{9}, 
T.~A.~Bell \inst{1}, 
M.~Emprechtinger \inst{1},
J.~H.~Black \inst{24},
G.~A.~Blake \inst{1},  
F.~Boulanger \inst{25}, 
S.~Cabrit \inst{32},
E.~Caux \inst{4,5},
C.~Ceccarelli \inst{6}
J.~Cernicharo \inst{7},
A.~Coutens \inst{4.5}, 
N.~R.~Crockett \inst{2},
F.~Daniel \inst{7,9},
E.~Dartois \inst{25},   
M.~De~Luca \inst{9},  
M.-L.~Dubernet \inst{10,11},
P.~Encrenaz \inst{9},
E.~Falgarone \inst{9}, 
T.~R.~Geballe \inst{27}, 
B.~Godard \inst{9},   
T.~F.~Giesen \inst{12},
J.~R.~Goicoechea \inst{7},
C.~Gry \inst{13},  
H.~Gupta\inst{13}, 
P.~Hennebelle \inst{9},  
P.~Hily-Blant \inst{6}, 
R.~Ko{\l}os \inst{28}, 
J.~Kre{\l}owski \inst{26}, 
C.~Joblin \inst{4,5},
D.~Johnstone \inst{15},
M.~Ka\'{z}mierczak \inst{26},  
S.~D.~Lord \inst{16},
S.~Maret \inst{6},
P.~G.~Martin \inst{17},
J.~Mart\'in-Pintado \inst{7},  
G.~J.~Melnick \inst{18},
K.~M.~Menten \inst{8},
R.~Monje \inst{1},  
B.~Mookerjea \inst{29},   
P.~Morris \inst{16},
J.~A.~Murphy  \inst{19},
V.~Ossenkopf \inst{12,20},
L.~Pagani \inst{32},
J.~C.~Pearson \inst{13},
M.~P\'erault \inst{9},
C.~Persson \inst{24},  
R.~Plume \inst{21},
S.-L.~Qin \inst{12},
M.~Salez \inst{32},  
S.~Schlemmer \inst{12},
M.~Schmidt \inst{30},  
P.~Sonnentrucker \inst{3},  
J.~Stutzki \inst{12},
D.~Teyssier \inst{31},  
N.~Trappe \inst{19},
F.~F.~S.~van der Tak \inst{20},
C.~Vastel \inst{4,5},
S.~Wang \inst{2}, 
H.~W.~Yorke \inst{13},
S.~Yu \inst{13},
J.~Zmuidzinas \inst{1},
A.~Boogert \inst{16}, 
N.~Erickson \inst{23},
A.~Karpov \inst{1}, 
J.~Kooi \inst{1},
F.~W.~Maiwald \inst{13}, 
R.~Schieder \inst{12},
P.~Zaal \inst{20}
}
\institute{California Institute of Technology, Cahill Center for Astronomy and Astrophysics 301-17, Pasadena, CA 91125 USA\\
             \email{dcl@caltech.edu}
\and Department of Astronomy, University of Michigan, 500 Church Street, Ann Arbor, MI 48109, USA 
\and  Department of Physics and Astronomy, Johns Hopkins University, 3400 North Charles Street, Baltimore, MD 21218, USA
\and Centre d'\'etude Spatiale des Rayonnements, Universit\'e de Toulouse [UPS], 31062 Toulouse Cedex 9, France
\and CNRS/INSU, UMR 5187, 9 avenue du Colonel Roche, 31028 Toulouse Cedex 4, France
\and Laboratoire d'Astrophysique de l'Observatoire de Grenoble, 
BP 53, 38041 Grenoble, Cedex 9, France.
\and Centro de Astrobiolog\'ia (CSIC/INTA), Laboratiorio de
Astrof\'isica Molecular, Ctra. de Torrej\'on a Ajalvir, km 4 28850,
Torrej\'on de Ardoz, Madrid, Spain
\and Max-Planck-Institut f\"ur Radioastronomie, Auf dem H\"ugel 69, 53121 Bonn, Germany 
\and LERMA, CNRS UMR8112, Observatoire de Paris and \'Ecole Normale Sup\'erieure, 24 Rue Lhomond, 75231 Paris Cedex 05, France
\and LPMAA, UMR7092, Universit\'e Pierre et Marie Curie,  Paris, France
\and  LUTH, UMR8102, Observatoire de Paris, Meudon, France
\and I. Physikalisches Institut, Universit\"at zu K\"oln,
              Z\"ulpicher Str. 77, 50937 K\"oln, Germany
\and Jet Propulsion Laboratory,  Caltech, Pasadena, CA 91109, USA
\and Departments of Physics, Astronomy and Chemistry, Ohio State University, Columbus, OH 43210, USA
\and National Research Council Canada, Herzberg Institute of Astrophysics, 5071 West Saanich Road, Victoria, BC V9E 2E7, Canada 
\and Infrared Processing and Analysis Center, California Institute of Technology, MS 100-22, Pasadena, CA 91125
\and Canadian Institute for Theoretical Astrophysics, University of Toronto, 60 St George St, Toronto, ON M5S 3H8, Canada
\and Harvard-Smithsonian Center for Astrophysics, 60 Garden Street, Cambridge MA 02138, USA
\and  National University of Ireland Maynooth. Ireland
\and SRON Netherlands Institute for Space Research, PO Box 800, 9700 AV, Groningen, The Netherlands
\and Department of Physics and Astronomy, University of Calgary, 2500
University Drive NW, Calgary, AB T2N 1N4, Canada
\and MPI f\"ur Sonnensystemforschung, D 37191 Katlenburg-Lindau,
Germany
\and University of Massachusetts, Astronomy Dept., 710 N. Pleasant
St., LGRT-619E, Amherst, MA 01003-9305  U.S.A 
\and Chalmers University of Technology, G\"oteborg, Sweden
\and Institut d'Astrophysique Spatiale (IAS), Orsay, France
\and Nicolaus Copernicus University, Torun, Poland
\and Gemini Telescope, Hilo, Hawaii, USA
\and Institute of Physical Chemistry, PAS, Warsaw, Poland
\and Tata Institute of Fundamental Research, Homi Bhabha Road, Mumbai 400005, India
\and Nicolaus Copernicus Astronomical Center, Poland
\and European Space Astronomy Centre, ESA, Madrid, Spain
\and LERMA \& UMR8112 du CNRS, Observatoire de
 Paris, 61, Av. de l'Observatoire, 75014 Paris, France}

\abstract{We present Herschel/HIFI observations of the fundamental
  rotational transitions of ortho- and para-H$_2^{16}$O and
  H$_2^{18}$O in absorption towards Sagittarius~B2(M) and W31C. The
  ortho/para ratio in water in the foreground clouds on the line of
  sight towards these bright continuum sources is generally consistent
  with the statistical high-temperature ratio of 3, within the
  observational uncertainties. However, somewhat unexpectedly, we
  derive a low ortho/para ratio of $2.35 \pm 0.35$, corresponding to a
  spin temperature of $\sim$27~K, towards Sagittarius~B2(M) at
  velocities of the expanding molecular ring. Water molecules in this
  region appear to have formed with, or relaxed to, an ortho/para
  ratio close to the value corresponding to the local temperature of
  the gas and dust.}

   \titlerunning{Ortho/para ratio in interstellar water}
	\authorrunning{Lis et al.}
   \maketitle
%

\section{Introduction}

Water molecules play an essential role in the physics and chemistry of
the dense interstellar medium (ISM). Water is one of the main
reservoirs of oxygen, and as an important coolant of dense gas it
strongly affects its star formation properties.  For a molecule with
two hydrogen atoms, such as water, the ortho/para ratio is temperature
dependent and, in principle, the temperature of the medium in which
the proton spin state populations last equilibrated can be deduced from the
observations. Water is an asymmetric top molecule, with energy levels
labeled by the set of quantum numbers $J$, $K_{-1}$, $K_{+1}$, where
$K_{-1}$, $K_{+1}$ are the limiting prolate and oblate symmetric top
quantum numbers.
Levels with $K_{-1}+K_{+1}$ even are para, and those with
$K_{-1}+K_{+1}$ odd are ortho. There are no fast radiative transitions
between the two species of water, ortho and para.  However, 
  a collision resulting in a proton exchange, with a proton or H$_3^+$
  ion, can cause an ortho-para conversion in a water molecule. Such
conversion can also result from interaction between a water molecule
and the grain surface on which it is adsorbed. The lowest ortho state
of water is 34.2~K above the para ground state. The ortho/para ratio
in water is 3 in the high temperature limit. Departures from this
limiting value are seen for nuclear spin temperatures below
$\sim$50~K, with the ratio dropping to 1.5 at $\sim$18~K (see Fig.~4
of \citealt{mumma87}).

The ground-state rotational transition of o-water at 557~GHz was
studied extensively in the ISM by the SWAS and Odin satellites
\citep{melnick05, hjalm04}. However, the HIFI instrument
\citep{graauw10} aboard the Herschel Space Observatory
\citep{pilbratt10} allows for the first time heterodyne studies of the
fundamental p-water line at 1113~GHz. Measurements of the
ortho/para ratio in water, by means of absorption spectroscopy in cold
foreground clouds on sightlines towards bright submillimeter continuum
sources, can provide key insights into the thermal history of the
gas. The ortho/para ratio in water has been measured in several
solar system comets and the derived nuclear spin temperatures are
typically of order 30~K (see \citealt{crovi97}; \citealt{kawakita04}).

\section{Observations}

HIFI observations of the ortho and para H$_2^{16}$O and H$_2^{18}$O
towards Sagittarius~B2(M) and W31C presented here were carried out
between 2010 March 1 and March 5, using the dual beam switch (DBS)
observing mode, as part of guaranteed time key programs \emph{HEXOS:
  Herschel/HIFI Observations of Extraordinary sources: The Orion and
  Sagittarius B2 Starforming Regions} and \emph{PRISMAS: Probing
  Interstellar Molecules with Absorption Line Studies}. The source
coordinates are: $\alpha_{J2000} = 17^{\rm h}47^{\rm m}20.35^{\rm s}$
and $\delta_{J2000} = -28^{\circ}23^{\prime}03.0^{\prime\prime}$ for
Sagittarius~B2(M) and $\alpha_{J2000} = 18^{\rm h}10^{\rm m}28.7^{\rm
  s}$ and $\delta_{J2000} = -19^{\circ}55^{\prime}50.0^{\prime\prime}$
for W31C. The DBS reference beams lie approximately 3$^{\prime}$ east
and west (i.e. perpendicular to the roughly north--south elongation of
Sagittarius~B2). We used the HIFI Wide Band Spectrometer (WBS)
providing a spectral resolution of 1.1~MHz ($\sim$0.6 km\,s$^{-1}$ at
557~GHz) over a 4~GHz IF bandwidth. The spectra presented here are
equally-weighted averages of the H and V polarizations, reduced using
HIPE (Ott 2010) with pipeline version 2.6. The resulting Level~2
double sideband (DSB) 
spectra were exported to the FITS format for a subsequent data
reduction and analysis using the IRAM GILDAS package.

The band 1a, 1b and 4b spectral scans of Sagittarius~B2(M) consist of 
DSB spectra with a redundancy of 8, which gives
observations of a specific lower or upper sideband frequency with 8 different
settings of the local oscillator (LO). This observing mode allows for
the deconvolution and isolation of a single sideband (SSB) spectrum
\citep{comito02}. We applied the standard deconvolution routine within
CLASS. 
The observations of water
lines in W31C were obtained using the DBS single point observing mode
with 3 shifted LO settings that were averaged to produce the final
spectra. The HIFI beam size at 557 and 1113~GHz is
38$^{\prime\prime}$ and 21$^{\prime\prime}$, respectively, with a main
beam efficiency equal to $\sim$0.68.

HIFI spectra towards strong continuum sources are affected by standing
waves. For sources with narrow lines, these standing waves can be
removed efficiently in the pipeline using the \emph{FitHifiFringe}
task in HIPE. However, in the case of Sagittarius~B2(M), the period of
standing waves is comparable to the line widths in the band 1 data.
Given the high density of lines in this source in the low-frequency
HIFI bands, the standing wave removal can lead to artifacts in the
resulting spectra. We, therefore, have not attempted to ``defringe''
these data. 

\section{Results}

Figure~1 shows spectra of the ground-state rotational transitions of
o- and p-H$_2^{16}$O and H$_2^{18}$O towards Sagittarius~B2(M). Lines
of H$_2^{16}$O are completely saturated over a wide range of
velocities. Absorption in the envelope of Sagittarius~B2 is seen around
62~km\,s$^{-1}$, 
while absorption at velocities between $-140$ and
$40$~km\,s$^{-1}$ is due to the foreground gas. 

\begin{figure}[tb]
   \centering
   \includegraphics[width=0.98\columnwidth]{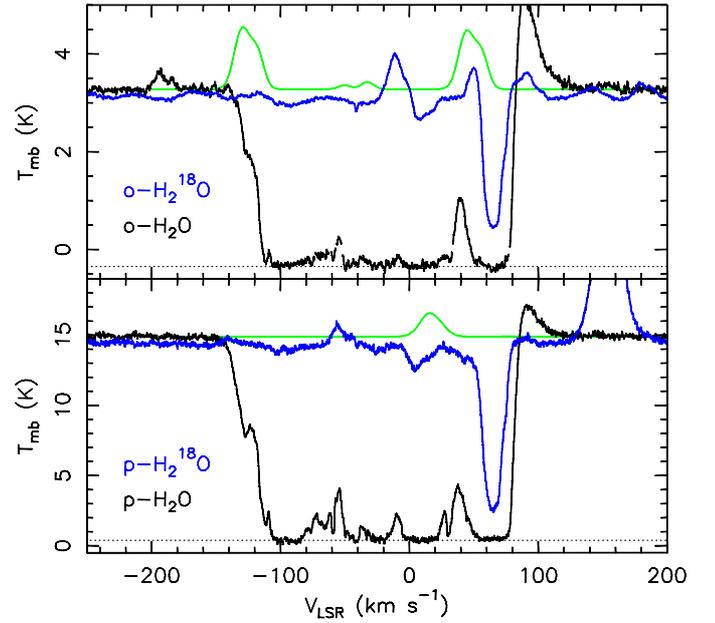}
  \caption{\emph{(Upper)} Spectra of the ground-state
     transitions of o-H$_2^{16}$O (band 1b) and H$_2^{18}$O (band
     1a) towards Sagittarius~B2(M) (black and blue histograms, respectively).
     \emph{(Lower)} Corresponding spectra of the
     p-H$_2^{16}$O and H$_2^{18}$O transitions (band 4b). Green
     lines show the model for SO$_2$ emission from the hot core
     described in the text.}
         \label{fig:spec}
\end{figure}

 The o-H$_2^{16}$O transition can be observed in HIFI
bands 1a and 1b. While the two observations give the same
line shape, the depth of the absorption is a few percent
above the zero continuum level in band 1a and a few percent below the
zero continuum level in band 1b (Fig.~1, upper panel). This effect is due
to small imbalances in the mixer sideband ratios. The sideband
deconvolution can, in principle, fit the sideband gains. However, the
H$_2$O line is near the band edge and can only observed in the upper
sideband in band 1a and only in the lower sideband in band 1b. The
resulting determination of the sideband gains is thus not completely
accurate. In the following analysis, we corrected for this
instrumental effect by assuming that the deepest absorption in both
H$_2^{16}$O lines defines the zero continuum level.

Sagittarius~B2(M) has a rich emission spectrum, which adds to the dust
continuum, and against which the foreground clouds can absorb. Some of
the strongest and most numerous emission lines seen in the spectrum
are those of SO$_2$, which we have modelled assuming LTE.\footnote{We
  made use of the myXCLASS program
  (http://www.astro.uni-koeln.de/projects/schilke/XCLASS), which
  accesses the CDMS (M\"uller at al. 2001, M\"uller et al. 2005;
  http://www.cdms.de) and JPL (Pickett et al. 1998;
  http://spec.jpl.nasa.gov) molecular databases.} The resulting model,
shown as green lines in Figure~1, includes two relatively strong
SO$_2$ lines within the o-H$_2$O spectrum, and a weaker line within
the p-H$_2$O spectrum. These lines are included, as an additional
background, in the calculation of the H$_2$O optical depth. Another
SO$_2$ line is present in the o-H$_2^{18}$O spectrum at
$-15$~km\,s$^{-1}$ (Fig.~2, upper panel) and has been similarly
modelled. The full source model of S.-L.~Qin (private comm.) suggests
that no other strong lines are present in our spectra at velocities
corresponding to the foreground clouds.

Figure~2 (upper and middle panels) shows the o- and p-H$_2^{16}$O and
H$_2^{18}$O spectra divided by the background emission, including dust
continuum and the SO$_2$ lines. The o-H$_2^{16}$O is an
equally-weighted average of the band 1a and 1b data. The H$_2^{16}$O
lines are saturated over a wide range of velocities and thus unusable
for a quantitative analysis. However, we have identified several
velocity ranges with moderate saturation levels, marked with thick
horizontal lines in Figure~2. These can be identified with the
expanding molecular ring ($<-50$~km\,s$^{-1}$), a transition between
the 4~kpc arm and the Orion arm ($-12$ to $-7$~km\,s$^{-1}$), the
Sagittarius arm (5 to 20~km\,s$^{-1}$), and the Scutum arm (27 to
35~km\,s$^{-1}$; possibly blended with the Sagittarius~B2 envelope;
see e.g. Neufeld et al. 2000). Assuming that the foreground absorption
completely covers the continuum and all water molecules are in the
ground state (a reasonable assumption for the diffuse foreground
clouds given the very high spontaneous emission rate coefficients for
the ground-state water lines, $3.458 \times 10^{-3}$ and $1.842
  \times 10^{-2}$~s$^{-1}$ for the ortho and para lines,
  respectively), we derive optical depths of the o- and p-water lines
($\tau = -\ln I/I_o$). The resulting optical depth ratio is shown in
Figure~2 (lower panel; left intensity
scale). 
An ortho/para optical depth ratio of 1 corresponds to a column density
ratio of 2. The resulting ortho/para column density ratio is given by
the right hand scale in Figure~2 (lower panel).

\begin{figure}[tb]
   \centering
   \includegraphics[width=0.98\columnwidth]{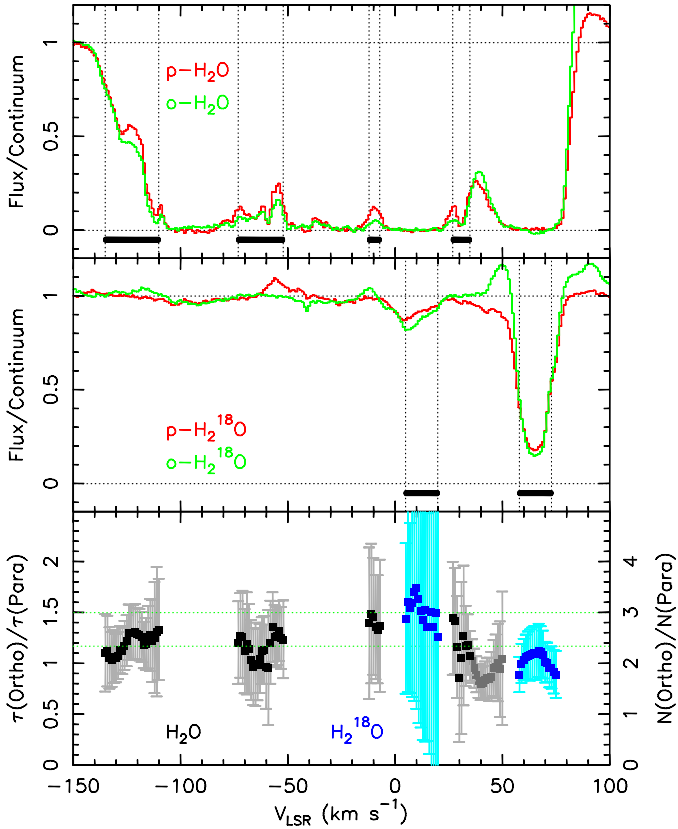}
  \caption{\emph{(Upper)} Spectra of the ground state o- and
     p-H$_2^{16}$O lines towards Sagittarius B2(M) normalized by the
     continuum (green and red histograms, respectively).
     \emph{(Middle)} Corresponding spectra of the H$_2^{18}$O lines.
     \emph{(Lower)} The ortho/para optical depth ratio (left
     scale) and column density ratio (right scale) as a function of
     LSR velocity. Black and blue points correspond to the H$_2^{16}$O
     and H$_2^{18}$O measurements, respectively. }
         \label{fig:sgrb2}
\end{figure}

The uncertainty in the line optical depth in a given velocity channel
is given by $\delta \tau = \exp (\tau) \times \delta I /I_o$. The
errorbars in Figure~2 (lower panel), are computed under the
conservative assumption $\delta I/I_o = 0.05$ (maximum uncertainty,
dominated by the standing waves, based on a comparison of the
continuum normalized spectra of the o-H$_2$O line in bands 1a and 1b;
the signal-to-noise ratio for the p-H$_2$O line is comparable, given
the stronger continuum, and the amplitude of the standing waves also
scales with the continuum strength). Relative uncertainties of the o-
and p-water optical depths have been added in quadrature.

\begin{figure}[tb]
   \centering
   \includegraphics[width=0.98\columnwidth]{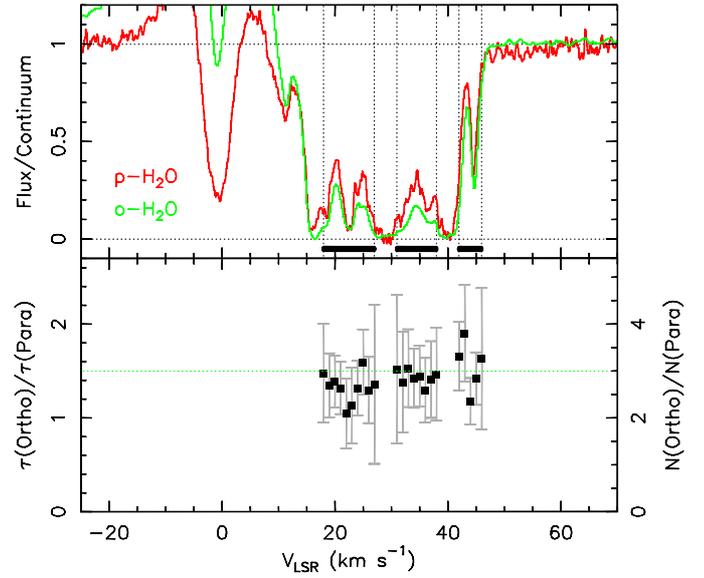}
  \caption{\emph{(Upper)} Spectra of the ground state o- and
     p-H$_2^{16}$O lines towards W31C normalized by the continuum
     (green and red histograms, respectively). \emph{(Lower)}
     The ortho/para optical depth ratio (left scale) and column
     density ratio (right scale) as a function of LSR velocity. }
         \label{fig:w31c}
\end{figure}

Table~1 gives weighted averages of the ortho/para column density ratio
in different velocity ranges towards Sagittarius~B2(M), along with the
corresponding uncertainties.  Since the individual measurements
  are not independent but dominated by instrumental systematics, we
  use a more conservative estimate of the uncertainty of the mean for
  the two velocity intervals corresponding to the expanding molecular
  ring, computed from the peak-peak variation between the individual
  data points. The H$_2$O (ortho+para) column densities are also
included, assuming an H$_2^{16}$O/H$_2^{18}$O ratio of 500 for the
5--20~km\,s$^{-1}$ component. We compute H$_2$ column densities in the
foreground gas using the method employed in Lis et al. (2001), based
on $^{13}$CO absorption data, assuming a CO abundance of $1\times
10^{-4}$ and a $^{12}$CO/$^{13}$CO ratio of 60 in the local gas in the
Sagittarius arm (5--20~km\,s$^{-1}$ velocity range) and 30 in the
remaining velocity intervals. The resulting column densities should be
accurate to within a factor of 2. The H$_2$O abundance in the various
components is generally consistent with that derived by Neufeld et al.
(2000; $4-7 \times 10^{-7}$). The derived H$_2$O abundance in the
5--20~km\,s$^{-1}$ component, based on the H$_2^{18}$O measurements,
is a factor of 3--6 higher than that derived for the other components
based on the H$_2^{16}$O data. However, the H$_2^{18}$O optical depth
in this velocity range is low leading to large uncertainties. The
water column density and abundance in this velocity range would be a
factor of 2 lower if the gas responsible for the absorption is located
in the Galactic center region (H$_2^{16}$O/H$_2^{18}$O ratio of 250)
rather than in the Sagittarius arms.

The ortho/para ratio in the two velocity ranges between $-12$ and
20~km\,s$^{-1}$ is consistent with the high-temperature limit of 3,
within the uncertainties. The ratio in the Scutum arm, at 27 to
35~km\,s$^{-1}$ is lower than 3. However, this component may be blended
with the Sagittarius~B2 envelope. The low ortho/para ratios derived at the
envelope velocities, from both H$_2^{16}$O and H$_2^{18}$O data, are
likely caused by the excitation effects. Given the higher density of
the gas, the assumption that all water molecules are in the ground
state is no longer correct. The column density of
o-H$_2$O is more strongly affected, resulting in a lower
apparent ortho/para ratio. 
In fact, we do see wings of the
o-H$_2^{18}$O line in emission at the envelope velocities (Fig.~2;
middle panel). 
Deriving the ortho/para ratio in water in the Sagittarius~B2
envelope requires detailed radiative transfer modelling and is beyond
the scope of the present Letter.

\begin{table}[tb]
\label{tab:spec}
\caption{Column densities (cm$^{-2}$) and ortho/para ratios towards Sagittarius~B2(M).}
\begin{tabular}{ccccc}
\hline \hline
\rule[-3mm]{0mm}{8mm}$V$ (km\,s$^{-1}$) & $O/P$ & $N$(H$_2$O)
& $N$(H$_2$) & $X$(H$_2$O)\\
\hline
$-135$ to $-110$& $2.35 \pm 0.3$ & $2.0 \times 10^{14}$ & $5 \times 
10^{20}$ & $4 \times 10^{-7}$ \\
$-73$   to $-52$  & $2.35 \pm 0.4$ & $3.7 \times 10^{14}$ & $6
\times 10^{20}$ & $6 \times 10^{-7}$\\ 
$-12$   to $-7$    & $2.8 \pm 0.5$   & $1.2 \times 10^{14}$ & $4
\times 10^{20}$ & $3\times 10^{-7}$\\
5           to 20        & $3.0 \pm 0.6$   & $6.7 \times 10^{15}$ &
$3 \times 10^{21}$ & $2 \times 10^{-6}$ \\
27         to 35        & $2.3 \pm 0.3$   & $1.7 \times 10^{14}$ &
$6 \times 10^{20}$ & $3 \times 10^{-7}$ \\
\hline
\end{tabular}
\end{table}

We derive a low ortho/para ratio, $2.35 \pm 0.35$, at velocities $<
-50$~km\,s$^{-1}$, corresponding to the expanding molecular ring. In
this case, the measurement uncertainties are small enough that the
ortho/para ratio of 3 appears to be ruled out by our data.

The o- and p-water spectra towards W31C, normalized by the continuum,
are shown in Figure~3 (upper panel). Weak p-H$_2^{18}$O absorption is
seen at the cloud systemic velocity, but no o- or p-H$_2^{18}$O
absorption is detected in the foreground clouds. Once again, we have
identified 3 velocity ranges where the o- and p-H$_2^{16}$O lines are
not completely saturated (thick horizontal lines in Fig.~3, upper
panel). The resulting ortho/para ratio is relatively uniform, $2.8 \pm
0.2$, consistent with the high-temperature ratio of 3 within the
measurement uncertainty (Fig.~3, lower panel).

\section{Discussion}

In the high-temperature limit one might expect an ortho/para ratio
close to 3, when water is first formed. In the gas-phase, the excess
energy of the exothermic reactions (e.g., recombination of H$_3$O$^+$)
could lead to spin equilibration. For water molecules desorbed from
ices on grain surfaces, one might also expect the initial ortho/para
ratio to be 3---if there is enough energy to desorb a water molecule,
either thermally or non-thermally, there is likely enough energy to
populate numerous rotational states of the gas-phase species, both
ortho and para, preserving a ratio close to 3 independent of the
ortho/para ratio in the ice. However, the excess energy of formation
is shared with the surface and the water ortho/para ratio may rapidly
equilibrate at the grain temperature (e.g., Limbach et al. 2006).

Once water molecules are in the gas phase, collisions with atomic and
molecular ions (H$^+$ and H$_3^+$) can lead to proton exchanges and,
over time, produce a gas-phase ortho/para ratio below 3 and
potentially commensurate with the gas temperature (see the discussion
of H$_2$ conversion by \citealt{flower06}). As the temperature
approaches zero, all water molecules will be in the ground para state
if collisions with ions can efficiently exchange the ortho and para
states. If there are no ions present, and therefore no barrierless
collisions to exchange the ortho and para states of water, the
ortho/para ratio of 3 will remain, independent of the temperature.
Since ions are present in the ISM, we can expect the water ortho/para
ratio to be generally lower than 3,  given sufficient time. The time
scale of the ortho/para conversion depends on the gas density. In
dense clouds, there is a greater likelihood that the ortho/para ratio
equilibrates at the gas kinetic temperature, as long as the density of
ions is sufficient. In warmer, diffuse clouds, equilibrium would favor
higher ortho/para ratios, but at lower densities equilibrium is less
likely.  Assuming a gas density of $10^4$~cm$^{-3}$, a fractional
  abundance of protonated ions of $10^{-8}$, and a rate coefficient
  of $10^{-9}$~cm$^{-3}$s$^{-1}$, we estimate the time scale of the
  ortho-para equilibration to be of order $3 \times 10^5$~years.
 
We derive a low ortho/para ratio of $2.35 \pm 0.35$, corresponding to a
spin temperature of $\sim$27~K, towards Sagittarius~B2(M) at
velocities corresponding to the expanding molecular ring. The low
ortho/para ratio may suggest water formation on dust grains, with
water molecule spin populations equilibrated at the dust temperature.
Alternatively, the high-temperature ratio of 3 may have relaxed in the
gas-phase to a value in line with the kinetic temperature of these
clouds---a gas/dust temperature of order 30~K is quite reasonable for
these relatively diffuse clouds within $\sim$200~pc from the Galactic
center (e.g., Tieftrunk et al. 1994). The water ortho/para ratio has
now been measured in atmospheres of 9 solar system comets. The derived
spin temperatures range between 23 and 34~K, with an average of
$\sim$29~K (J.~Crovisier, priv. comm.), close to our measurement in
the expanding molecular ring.

The higher ortho/para ratio, consistent with the statistical ratio of
3, that we derive towards Sagittarius~B2(M) in the velocity range
$-12$ to 20~km\,s$^{-1}$ may indicate higher gas or dust temperatures
(e.g., Gardner 1988). Schilke et al. (2010) also derive a higher
H$_2$O$^+$ spin temperature towards this source at positive
velocities, as compared to the line-of-sight clouds at negative
velocities. In addition, strong chloronium absorption has been
detected towards Sagittarius~B2(S) at velocities between $-10$ and
20~km\,s$^{-1}$ \citep{lis10}. Since the H$_2$Cl$^+$ abundance is
enhanced in warm, UV-irradiated regions, this indicates the presence
of a warm, diffuse gas component in this velocity range.

In summary, in the diffuse regions studied here, both water formation
mechanisms (gas and solid state) can contribute. With sufficient time
(or higher densities with subsequently faster gas-phase reactions)
chemistry in the gas can 
drive the spin temperature towards the gas temperature. In addition,
it has been postulated (Hollenbach et al. 2009) that photodesorption
of water ices in low-density environments can release frozen molecules
with a spin temperature that coincides with that of the dust grains.
If the gas and dust temperatures are not equal this can lead to a
ratio that reflects a mixture of these effects. Modeling the
ortho/para ratio evolution is beyond the scope of this Letter, but the
theme is that a ratio below 3 reflects cold environments (gas or dust
temperatures below about 50~K).
 
With improved calibration and better understanding of the instrumental
effects, more accurate determination of the water ortho/para ratio in
these and other sources will be possible in the future. However, the
present work clearly demonstrates the outstanding spectroscopic
capabilities of HIFI for providing robust constraints for the physical
conditions and chemistry of the ISM.

\begin{acknowledgements}
  HIFI has been designed and built by a consortium of institutes and
  university departments from across Europe, Canada and the United
  States under the leadership of SRON Netherlands Institute for Space
  Research, Groningen, The Netherlands and with major contributions
  from Germany, France and the US. Consortium members are: Canada:
  CSA, U.Waterloo; France: CESR, LAB, LERMA, IRAM; Germany: KOSMA,
  MPIfR, MPS; Ireland, NUI Maynooth; Italy: ASI, IFSI-INAF,
  Osservatorio Astrofisico di Arcetri-INAF; Netherlands: SRON, TUD;
  Poland: CAMK, CBK; Spain: Observatorio Astronomico Nacional (IGN),
  Centro de Astrobiolog\'ia (CSIC-INTA). Sweden: Chalmers University
  of Technology-MC2, RSS \& GARD; Onsala Space Observatory; Swedish
  National Space Board, Stockholm University - Stockholm Observatory;
  Switzerland: ETH Zurich, FHNW; USA: Caltech, JPL, NHSC. Support for
  this work was provided by NASA through an award issued by
  JPL/Caltech. D.~C.~L. is supported by the NSF, award AST-0540882 to
  the CSO. A portion of this research was performed at the Jet
  Propulsion Laboratory, California Institute of Technology, under
  contract with the National Aeronautics and Space Administration.
\end{acknowledgements}


\begin{thebibliography}{}


\bibitem[Comito \& Schilke(2002)]{comito02} Comito, C., \& Schilke,
  P.\ 2002, \aap, 395, 357

\bibitem[Crovisier et al.(1997)]{crovi97} Crovisier, J., Leech, K.,
  Bockel\'{e}e-Morvan, D., et al. 1997, Science, 275, 1904

\bibitem[Flower et al.(2006)]{flower06} Flower, D., Pineau des
  For\^{e}ts, G., \& Walmsley, C.~M. 2006, A\&A, 449, 621

\bibitem[Gardner et al. (1988)]{gardner98} Gardner, F.~F., Boes, F.,
  \& Winnewisser, G. 1988, A\&A, 196, 207

\bibitem[de Graauw(2010)]{graauw10} de Graauw, Th.\ et al.\, 2010,
  A\&A, in press 

\bibitem[Greaves et al.(1996)]{greaves96} Greaves, J. S. \& Nyman,
  L.-A. 1996, A\&A, 305, 950

\bibitem[Hjalmarson(2004)]{hjalm04} Hjalmarson, \AA\ 2004, ASR, 34, 504

\bibitem[Hollenbach et al.(2009)]{hollen09} Hollenbach, D., Kaufman,
    et al. 2009, ApJ, 690, 1497

\bibitem[Kawakita et al.(2004)]{kawakita04} Kawakita, H., Watanabe,
  J., et al. 2004, ApJ, 601, 1152

\bibitem[Limbach et al.(2006)]{limbach06} Limbach, H.-H., Buntkowsky,
  G., et al. 2006, ChemPhysChem, 7, 551

\bibitem[Lis et al. (2001)]{lis01} Lis, D.~C., Keene, J., Phillips,
  T.G., et al. 2001, ApJ, 561, 823

\bibitem[Lis et al.(2010)]{lis10} Lis, D.~C., Pearson, J.~C., Neufeld,
  D.~.A., et al. 2010, A\&A, this volume

\bibitem[Melnick \& Bergin(2005)]{melnick05} Melnick, G.~J., \&
  Bergin, E.~A. 2005, ASR, 36, 1027

\bibitem[M{\"u}ller et al.(2001)]{muller01} M{\"u}ller, H.~S.~P.,
  Thorwirth, S., et al. 2001, A\&A, 370, L49

\bibitem[Muller(2005)]{muller05} M{\"u}ller, H.~S.~P., Schl{\"o}der,
  et al. G. 2005, J. Mol. Struct., 742, 215

\bibitem[Mumma et al.(1987)]{mumma87} Mumma, M., Weaver, H.~A., \&
  Larson, H.~P. 1987, A\&A, 187, 419

\bibitem[Neufeld et al. (2000)]{neufeld00} Neufeld, D.~A., Ashby,
  M.~L.~N., Bergin, E.~A., et al. 2000, ApJ, 539, L111

\bibitem[Ott(2010)]{ott10} Ott, S. 2010, in ASP Conference Series,
  Astronomical Data Analysis Software and Systems XIX, Y. Mizumoto,
  K.-I. Morita, and M. Ohishi, eds., in press

\bibitem[Pickett et al.(1998)]{picket98} Pickett, H.~M., Poynter,
  R.~L., Cohen, E.~A., et al. 1998, J. Quant. Spectrosc. Radiat.
  Transfer, 60, 883

\bibitem[Pilbratt et al.(2010)]{pilbratt10} Pilbratt, G., et al.
  2010, A\&A, in press

\bibitem[Schilke et al. (2010)]{schilke10} Schilke, P., Comito, C. et
  al. 2010, this volume

\bibitem[Tieftrunk et al.(1994)]{tief94} Tieftrunk, A., Pineau des
  For\^{e}ts, G. et al. 1994, A\&A, 289, 579


\end{thebibliography}
\end{document}